\documentclass[aps,preprint, tightenlines,superscriptaddress,showpacs]{revtex4}

\usepackage{graphicx} 
\usepackage{dcolumn}  

\begin{document}

\vspace*{-3\baselineskip}
\resizebox{!}{3cm}{\includegraphics{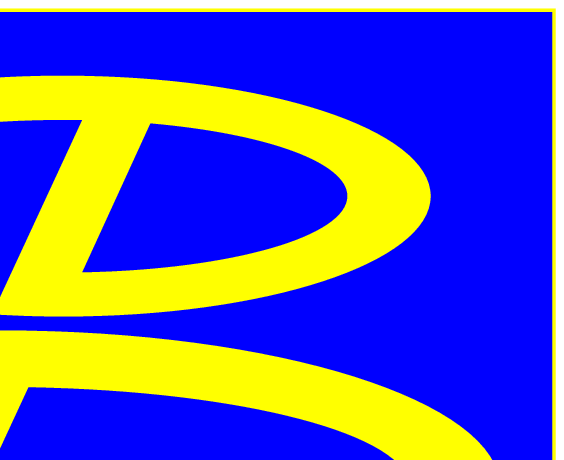}}

\preprint{\vbox{ 
                 \hbox{Belle Preprint 2007-29}
                 \hbox{KEK Preprint 2007-26}
}}

\title{
Study of charmonia in four-meson final states\\ 
produced in two-photon collisions}

\affiliation{Budker Institute of Nuclear Physics, Novosibirsk}
\affiliation{Chiba University, Chiba}
\affiliation{University of Cincinnati, Cincinnati, Ohio 45221}
\affiliation{Department of Physics, Fu Jen Catholic University, Taipei}
\affiliation{Justus-Liebig-Universit\"at Gie\ss{}en, Gie\ss{}en}
\affiliation{The Graduate University for Advanced Studies, Hayama}
\affiliation{Gyeongsang National University, Chinju}
\affiliation{Hanyang University, Seoul}
\affiliation{University of Hawaii, Honolulu, Hawaii 96822}
\affiliation{High Energy Accelerator Research Organization (KEK), Tsukuba}
\affiliation{Hiroshima Institute of Technology, Hiroshima}
\affiliation{University of Illinois at Urbana-Champaign, Urbana, Illinois 61801}
\affiliation{Institute of High Energy Physics, Chinese Academy of Sciences, Beijing}
\affiliation{Institute of High Energy Physics, Vienna}
\affiliation{Institute of High Energy Physics, Protvino}
\affiliation{Institute for Theoretical and Experimental Physics, Moscow}
\affiliation{J. Stefan Institute, Ljubljana}
\affiliation{Kanagawa University, Yokohama}
\affiliation{Korea University, Seoul}
\affiliation{Kyoto University, Kyoto}
\affiliation{Kyungpook National University, Taegu}
\affiliation{Swiss Federal Institute of Technology of Lausanne, EPFL, Lausanne}
\affiliation{University of Ljubljana, Ljubljana}
\affiliation{University of Maribor, Maribor}
\affiliation{University of Melbourne, School of Physics, Victoria 3010}
\affiliation{Nagoya University, Nagoya}
\affiliation{Nara Women's University, Nara}
\affiliation{National Central University, Chung-li}
\affiliation{National United University, Miao Li}
\affiliation{Department of Physics, National Taiwan University, Taipei}
\affiliation{H. Niewodniczanski Institute of Nuclear Physics, Krakow}
\affiliation{Nippon Dental University, Niigata}
\affiliation{Niigata University, Niigata}
\affiliation{University of Nova Gorica, Nova Gorica}
\affiliation{Osaka City University, Osaka}
\affiliation{Osaka University, Osaka}
\affiliation{Panjab University, Chandigarh}
\affiliation{Peking University, Beijing}
\affiliation{University of Pittsburgh, Pittsburgh, Pennsylvania 15260}
\affiliation{Princeton University, Princeton, New Jersey 08544}
\affiliation{RIKEN BNL Research Center, Upton, New York 11973}
\affiliation{Saga University, Saga}
\affiliation{University of Science and Technology of China, Hefei}
\affiliation{Seoul National University, Seoul}
\affiliation{Shinshu University, Nagano}
\affiliation{Sungkyunkwan University, Suwon}
\affiliation{University of Sydney, Sydney, New South Wales}
\affiliation{Tata Institute of Fundamental Research, Mumbai}
\affiliation{Toho University, Funabashi}
\affiliation{Tohoku Gakuin University, Tagajo}
\affiliation{Tohoku University, Sendai}
\affiliation{Department of Physics, University of Tokyo, Tokyo}
\affiliation{Tokyo Institute of Technology, Tokyo}
\affiliation{Tokyo Metropolitan University, Tokyo}
\affiliation{Tokyo University of Agriculture and Technology, Tokyo}
\affiliation{Toyama National College of Maritime Technology, Toyama}
\affiliation{Virginia Polytechnic Institute and State University, Blacksburg, Virginia 24061}
\affiliation{Yonsei University, Seoul}

  \author{S.~Uehara}\affiliation{High Energy Accelerator Research Organization (KEK), Tsukuba} 
  \author{I.~Adachi}\affiliation{High Energy Accelerator Research Organization (KEK), Tsukuba} 
  \author{H.~Aihara}\affiliation{Department of Physics, University of Tokyo, Tokyo} 
  \author{V.~Aulchenko}\affiliation{Budker Institute of Nuclear Physics, Novosibirsk} 
  \author{T.~Aushev}\affiliation{Swiss Federal Institute of Technology of Lausanne, EPFL, Lausanne}\affiliation{Institute for Theoretical and Experimental Physics, Moscow} 
  \author{A.~M.~Bakich}\affiliation{University of Sydney, Sydney, New South Wales} 
  \author{V.~Balagura}\affiliation{Institute for Theoretical and Experimental Physics, Moscow} 
  \author{E.~Barberio}\affiliation{University of Melbourne, School of Physics, Victoria 3010} 
  \author{A.~Bay}\affiliation{Swiss Federal Institute of Technology of Lausanne, EPFL, Lausanne} 
  \author{K.~Belous}\affiliation{Institute of High Energy Physics, Protvino} 
% ...
\collaboration{The Belle Collaboration}

\vspace*{10mm}
\begin{abstract}
We report measurements of charmonia produced in two-photon 
collisions and decaying to four-meson final states, where 
the meson is  either a
charged pion or a charged kaon. The analysis is based on a 
$395$~fb$^{-1}$ data sample accumulated with the Belle detector at the KEKB
electron-positron collider. We observe signals for the three $C$-even 
charmonia $\eta_c(1S)$, $\chi_{c0}(1P)$ and $\chi_{c2}(1P)$ 
in the $\pi^+\pi^-\pi^+\pi^-$, $K^+K^-\pi^+\pi^-$ and 
$K^+K^-K^+K^-$ decay modes. No clear signals for
$\eta_c(2S)$ production are found in these decay modes.
We have also studied resonant structures in charmonium decays
to two-body intermediate meson resonances. 
We report the products of the two-photon decay width
and the branching fractions, $\Gamma_{\gamma\gamma}{\cal B}$,
for each of the charmonium decay modes.
\end{abstract}


\maketitle

\tighten

\setcounter{footnote}{0}
\section{Introduction}
   The two-photon decay widths ($\Gamma_{\gamma\gamma}$)
of charge-conjugation ($C$)-even charmonium states
are important observables that are sensitive to the properties
of $c\bar{c}$ quarks inside charmonium bound states.
Precise measurements of widths can give valuable 
constraints on the models that describe the nature of heavy 
quarkonia~[1-3].

High luminosity electron-positron colliders are well suited 
to measurements of the two-photon production of 
charmonium states, since they provide a large flux of quasi-real
photons colliding at two-photon center-of-mass energies 
covering a wide range. Various charmonium states, including the
$\eta_c(1S)$, $\chi_{c0}$, $\chi_{c2}$, $\eta_c(2S)$ and
$\chi_{c2}(2P)$, have been observed so far in two-photon 
production processes.

In  resonance formation,
the production cross section from quasi-real two-photon
collisions at an $e^+e^-$ collider is 
proportional to the product of the two-photon decay width
of the resonance and 
its branching fraction (${\cal B}$) to the observed
final state: 
\[
\sigma\Big(e^+e^- \to (e^+e^-)R,\ R \to {\rm final~state}\Big)
\propto \Gamma_{\gamma\gamma}(R){\cal B}(R \to {\rm final~state}),
\]
where $R$ is the resonance.  For the above 
charmonium states, the two-photon decay widths
are mainly determined from two-photon collision measurements performed
in a small number of specific decay modes only since 
the production rate of the corresponding states
is limited.  The results have
rather large statistical errors and in some cases even larger systematic errors
originating from uncertainties in the corresponding charmonium 
branching fraction
to the observed final state. It is necessary
to reduce these errors with high-statistics measurements
in various decay channels. Through these measurements, we can also 
obtain ratios of branching fractions for various
charmonium decay modes.

  In this paper, we report the measurements of production of the 
charmonium states $\eta_c(1S)$, $\chi_{c0}$ and  $\chi_{c2}$
in two-photon collisions 
from their decays to four-meson final states $\pi^+\pi^-\pi^+\pi^-$, 
$K^+K^-\pi^+\pi^-$ and
$K^+K^-K^+K^-$, with the Belle 
detector. We have studied intermediate meson-resonant structures 
in these decay modes.  We have also searched for
production of the $\eta_c(2S)$ in these final states. 
Previous measurements of the same processes are found in Refs.~[4-10].
Belle previously reported measurements of two-photon 
production of these charmonia in several decay modes:
$\chi_{c2} \to \gamma J/\psi$~\cite{bellechic2}, $\chi_{c0}$ and
$\chi_{c2}$ decays to the two-meson 
final states $\pi^+\pi^-$ and $K^+K^-$~\cite{nkzw}
as well as $K^0_SK^0_S$~\cite{wtchen},
and $\eta_c(1S) \to p\bar{p}$~\cite{cckuo}.
Production of the $\eta_c(2S)$ in a two-photon process was observed 
in the $K^0_S K^{\mp}\pi^{\pm}$ mode by the CLEO and BaBar
experiments~\cite{cleo2s,babar}. This is the only decay
mode so far directly observed for this charmonium state. 

\section{Experiment and event selection}
\subsection{Data and the Belle detector}

We use data that corresponds to an integrated luminosity of 
395~fb$^{-1}$ recorded with
the Belle detector at the KEKB asymmetric-energy $e^+e^-$ collider~\cite{kekb}.
Since the beam energy dependence of 
two-photon processes is very small, we combine the on-resonance and
off-resonance data samples; the off-resonance data were taken 
60~MeV below the $\Upsilon(4S)$ $(\sqrt{s} = 10.58~{\rm GeV})$.
The analysis is made in the ``zero-tag'' mode, where  
neither the recoil electron nor positron is detected. We
restrict the virtuality of the incident photons to be small 
by imposing a
strict transverse-momentum  balance along the beam axis  
for the final-state hadronic system.

  A comprehensive description of the Belle detector is
given elsewhere~\cite{belle}. We mention here only the
detector components essential to the present measurement.
Charged tracks are reconstructed from hit information in a central
drift chamber (CDC) located in a uniform 1.5~T solenoidal magnetic field.
The $z$ axis of the detector and the solenoid are along the positron beam
direction,
with the positrons moving in the $-z$ direction.  The CDC measures the
longitudinal and transverse momentum components (along the $z$ axis and
in the $r\varphi$ plane, respectively).
Track trajectory coordinates near the
collision point are provided by a
silicon vertex detector (SVD).  Photon detection and
energy measurements are performed with a CsI(Tl) electromagnetic
calorimeter (ECL).
Kaons are identified using information from the
time-of-flight counters (TOF) and silica-aerogel Cherenkov
counters (ACC).  The ACC 
provides good separation between kaons and pions or muons at momenta above
1.2~GeV/$c$.  The TOF system consists of a barrel of 128 plastic scintillation
counters, and is effective in $K/\pi$ 
separation mainly for tracks with momentum below 1.2~GeV/$c$. 
Lower energy kaons are also identified using specific
ionization ($dE/dx$) measurements in the CDC.
The magnet return yoke is 
instrumented to form the $K_L$ and muon detector (KLM), 
which detects muon tracks and provides trigger signals.

Signal events are efficiently triggered by several kinds of 
track-triggers that require two or more CDC tracks with combinations
of TOF hits, ECL clusters or summed energies of ECL clusters.
No additional cuts on the trigger information are applied. 
The trigger conditions are complementary 
for the detection of four-prong events; we obtain an overall trigger
efficiency of $\sim$ 95\%.

 $K/\pi$ separation uses a likelihood ratio formed
from ACC, TOF and CDC information. 
No explicit lepton identification requirement is applied, since leptons are
not a large background source in exclusive four-prong events.


\subsection{Event selection}

Candidate events are selected as follows;
all variables in selection criteria (1)-(10) are measured in 
the laboratory frame.
(1) There are exactly four charged particles 
and each satisfies the following criteria:
$p_t > 0.1$~GeV/$c$, $dr < 5$~cm, $|dz| < 5$~cm,  
where $p_t$ is the transverse momentum of a track with respect
to the $z$ axis, and $dr$ and $dz$ are the radial and axial distances,
respectively, of the closest
approach (as seen in the $r\varphi$ plane) to the nominal collision point;
(2) the net charge of the four tracks is zero;
(3) at least two of the four tracks satisfy the following additional
criteria: $p_t > 0.4$~GeV/$c$, $dr < 1$~cm, and 
$-0.8660 < \cos \theta < +0.9563$,
where $\theta$ is the polar angle;
in addition, the following two criteria are applied 
for the four-track systems: 
(4) the scalar sum of the momenta ($\sum_i |p_i|$) of the four tracks 
must be smaller than 6~GeV/$c$; (5) the average of the $dz$ 
values for the four tracks is within 3~cm of the nominal collision point.

  In order to reject backgrounds from single-photon annihilation
or radiative events, the following criteria are required:
(6) the invariant mass of track combinations that satisfy   
criterion (3) should be smaller than 4.5~GeV/$c^2$ (here, zero mass
is assigned to each of the charged tracks), and (7) the missing mass squared
of the recoil against the track combination is larger than 2~(GeV/$c^2)^2$. 

 Calorimetry requirements are also applied:
(8) the total calorimeter energy in an event is smaller than 6~GeV; (9)
there is no electromagnetic cluster unassociated with a track 
({\it i.e.}, a neutral cluster)
whose energy exceeds 0.4~GeV; (10) there is 
no $\pi^0$ candidate whose transverse momentum is larger than 0.1~GeV/$c$.

 After all these selection criteria are applied, the four tracks are 
transformed into
the $e^+e^-$ center-of-mass (c.m.) frame, and the vector sum of their
transverse momenta with respect to the beam direction in the 
c.m.\ frame, $\Sigma {\bf p}^*_t$, is calculated. The variable 
$\Sigma {\bf p}^*_t$
approximates the transverse momentum of the two-photon-collision system.
We require (11) $|\Sigma {\bf p}^*_t|<0.1$~GeV/$c$ in order to enhance
the number of events from quasi-real two-photon collisions.  Figure~1 shows the
transverse momentum ($|\Sigma {\bf p}^*_t|$) distribution near the
selection region. 

\begin{figure}
\centering
\includegraphics[width=8cm]{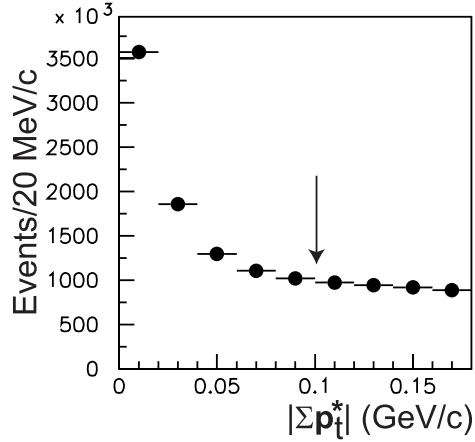}
\label{fig1}
\centering
\caption{Distribution of the vector sum of transverse momenta. 
Events to the right of the arrow are discarded.}
\end{figure}

\subsection{$K/\pi$ separation}

We apply particle identification requirements for each track. A likelihood  
ratio, ${\cal R} = L_K/( L_K + L_\pi)$, is calculated, 
where the kaon and pion likelihoods,  $L_K$ 
and $L_\pi$, are determined from information 
provided by the ACC, TOF and the CDC $dE/dx$ system. 
All tracks with ${\cal R} > 0.8$ are assigned to be kaons.
This requirement gives a typical identification efficiency of 90\% with a
probability of 3\% for a pion to be misidentified as a kaon.
All other tracks are treated as pions. 

We take (12) only the combinations in which
the net strangeness is conserved, $\pi^+\pi^-\pi^+\pi^-$, $K^+K^-\pi^+\pi^-$
and $K^+K^-K^+K^-$. These final-state combinations are sometimes
denoted $4\pi$, $2K2\pi$ and $4K$, respectively, in this paper.

When a pair of pions~(kaons) in the four prong decay of
a certain charmonium are misidentified as a pair of kaons~(pions),
they could produce a broad enhancement in the spectrum of the
misidentified channel at a mass very different
from that of the original charmonium.  However, these 
backgrounds do not make any
pronounced peaks near the three charmonia masses. 


\subsection{Rejection of ISR events}
  Two-photon events with relatively high $\gamma\gamma$ c.m.
energies (above 3.2~GeV in the case of Belle) are contaminated by 
background processes where a 
real or virtual photon is emitted in the direction of the incident electron. 
They are called ISR events or pseudo-Compton events,
in case the positron collides with 
the electron ($e^+e^- \to (e^+e^{-*})\gamma \to
{\rm hadrons}+\gamma$) or with the virtual photon
($e^+e^- \to (e^+\gamma^*)e^- \to {\rm hadrons}+e^+e^-$), respectively.
Such events have a kinematical correlation between
the $p_z$-component in the laboratory frame ($\Sigma p_z$) and 
the invariant mass of the final-state hadron system ($W$). 
We reject them with the following additional requirement: (13)
$\Sigma p_z > (W^2 - 49~{\rm GeV^2}/c^4 )/(14~{\rm GeV}/c^3) + 0.6~{\rm GeV}/c$.
These backgrounds are not harmful in the measurements
of $C$-even charmonia, since at lowest order 
they produce only $C$-odd hadron systems.  
We show a two-dimensional plot for $\Sigma p_z$ vs invariant mass
of the hadronic system for the four-pion final state candidate in Fig.~2. 

 After all these selections, we obtain  $6.37 \times 10^6$ $4\pi$ events,
$4.17 \times 10^5$ $2K2\pi$ events and 6003 $4K$ events in the four-meson
invariant mass region between 2.0~GeV/$c^2$ and 5.0~GeV/$c^2$.

\begin{figure}
\centering
\includegraphics[width=8cm]{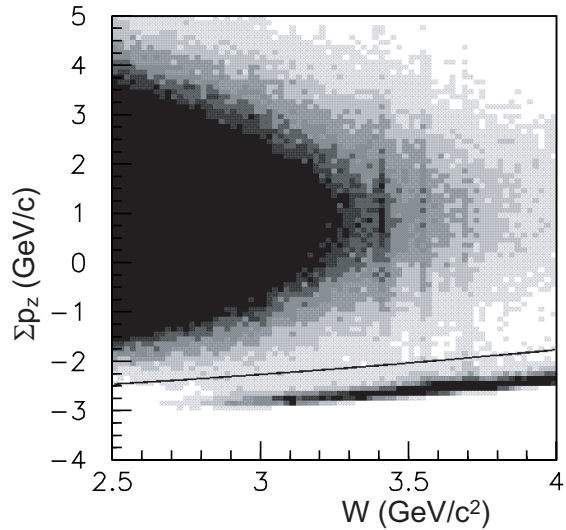}
\label{fig1}
\centering
\caption{ A two-dimensional plot for $\Sigma p_z$ vs invariant mass
of the hadronic system for the four-pion final state candidate. 
The curve is the cut criterion (13) described in the text.}
\end{figure}

\section{Yields for Charmonium Production}
\subsection{Yields of the charmonia}
  We measure the charmonium yields  
from the invariant mass distributions
in each of the three final-state processes. The distributions 
are shown in Fig.~3. Peaks at the $C$-even charmonium masses are 
assumed to be from two-photon production
processes if the transverse-momentum distribution peaks 
in the vicinity of $|\Sigma {\bf p}^*_t|= 0$. We discuss
the $|\Sigma {\bf p}^*_t|$ distributions in Sect.~V.C.

  We find clear enhancements for the 
$\eta_c(1S)$ (denoted hereafter as ``$\eta_c$'')
at $\sim 2.98$~GeV/$c^2$, $\chi_{c0}$ 
at $\sim 3.41$~GeV/$c^2$, and $\chi_{c2}$ at
$\sim 3.555$~GeV/$c^2$ in all the final states. 
We do not find any clear $\eta_c(2S)$ signals, which would
appear at around 3.63-3.65~GeV/$c^2$ in the invariant mass
distributions. A small peak near 3.69~GeV/$c^2$
in the $4\pi$ final state is from $\psi(2S) \to J/\psi \pi^+\pi^-$, $J/\psi
\to l^+l^-$, where the leptons are not identified and are treated as pions.
These events are attributed primarily to the double ISR
$e^+e^-$ annihilation process, 
$e^+e^- \to \psi(2S)\gamma\gamma$.  Upper limits for $\eta_c(2S)$ 
production will be derived in Sect.~VII.

The invariant-mass distribution in the vicinity of each 
charmonium peak is fitted to 
the sum of charmonium and background components using
$\chi^2$ minimization. We use a second-order polynomial for the
background component. The fitted regions are 
2.8-3.2~GeV/$c^2$, 3.3-3.5~GeV/$c^2$ and 3.5-3.6~GeV/$c^2$
for the analyses of the $\eta_c$, $\chi_{c0}$ and $\chi_{c2}$,
respectively.  The energy dependence of the efficiency
is small within each resonance region and is neglected.
We adopt the relativistic Breit-Wigner function for the
$\eta_c$ and $\chi_{c0}$ signals, which are smeared with a 
Gaussian function whose width is 
fixed to the invariant-mass resolution (ranges in 7-10~MeV/$c^2$ 
depending on the decay process) estimated by the signal Monte Carlo (MC)
simulation. 
In the fit to the $\chi_{c2}$, we use a simple Gaussian function 
with a floating width, because the natural
width is considerably smaller than the mass resolution.
The mass resolutions from the fits,  8 - 9 MeV/$c^2$, are consistent 
with the MC expectations at the $\chi_{c2}$ mass. 

The fit results are shown in Fig.~4 and are summarized in 
Table~\ref{tab:table1}. In the table, the masses are corrected
for the effects of a systematic shift seen in the signal MC mainly
due to the tails of kaon energy loss in the innermost detector region.
The correction size is 
typically $n_K \times (1\hbox{ - }2$\,MeV/$c^2$), where
$n_K$ is the number of kaons.
The systematic errors for the masses include the uncertainty due to
this effect (a half of the correction size) as well as the uncertainty
of the mass scale (2.0~MeV/$c^2$) originating from the uncertainty of
the momentum scale, while those for the widths are obtained by 
changing the mass resolution by 1~MeV/$c^2$. We confirm that the mass
and width of the $\eta_c$ are robust for the change of the fit region
as described in Sect.~V.B. 

We find a hint of a possible $J/\psi$ contribution from
the double ISR background in the $2K2\pi$ distribution.
This could affect the yield and resonance parameters of the $\eta_c$
determined from the fit. We have tried a fit that includes
a peak near the $J/\psi$ mass. The changes
of the $\eta_c$ yield (7\%) and the total width (1.9~MeV) are taken
into account in the systematic errors. There is no detectable
shift in the mass.
The fit including a $J/\psi$ peak gives the $J/\psi$ mass, 
$3093.6 \pm 2.0 \pm 2.4$~MeV/$c^2$,
where we apply the same correction and systematic error as applied for
$\eta_c \to  2K2\pi$. This provides a confirmation of 
the mass scale in the present measurement.   

Possible interference of the charmonia with the continuum 
component is not taken into account
in the present analysis.  It is difficult to separate the 
continuum-component amplitudes that interfere with the 
charmonium amplitude in each process, because different 
partial-wave components of the continuum appear within the 
background.  In general, the interference effects could 
give different mass/width fit results for each decay mode. 
However, for the three considered processes the values obtained for
these parameters from the no-interference fits to the separate decay modes
are consistent with each other.
The mass values for the $\eta_c$ are systematically higher than 
the world average value, $2979.8 \pm 1.2$~MeV/$c^2$~\cite{pdg}. 
The $\eta_c$ width and the parameters of $\chi_{c0}$ and $\chi_{c2}$
are consistent with the previous measurements. Table~\ref{tab:table1} also
shows the combined results from the measurements of the three decay processes,
where the central values are derived from the weighted means
of the three fits 
and the systematic errors are evaluated considering
the full correlations between the error sources.

\begin{figure}
\centering
\includegraphics[width=12cm]{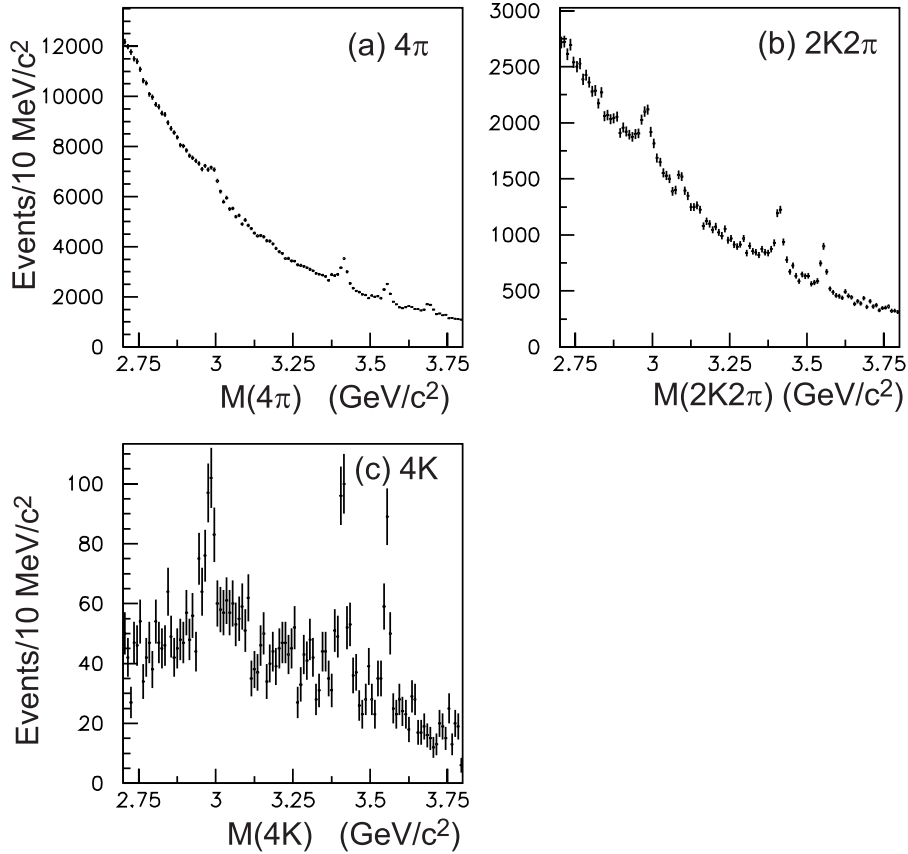}
\label{fig3}
\centering
\caption{Invariant mass distributions for the four-meson
final states, 
(a)$4\pi$, (b)$2K2\pi$ and (c)$4K$.}
\end{figure}

\begin{figure}
\centering
\includegraphics[width=12cm]{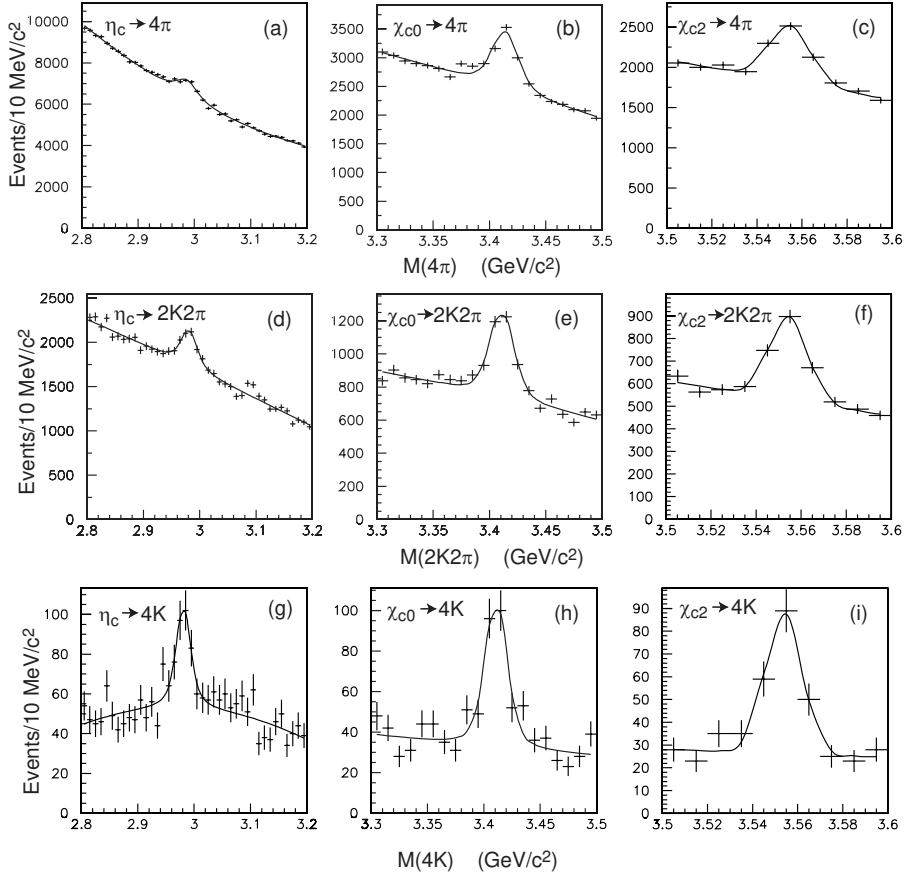}
\label{fig4}
\centering
\caption{Results of the fits to the invariant mass distributions 
for (a,b,c)~$4\pi$, (d,e,f)~$2K2\pi$ and (g,h,i)~$4K$  final states
in the charmonium mass regions.
}
\end{figure}

\begin{table}
\caption{Results for the resonance parameters obtained from  the
fits to the invariant-mass distributions. The third column
shows the Breit-Wigner width for the $\eta_c$ and $\chi_{c0}$, 
and the Gaussian standard deviation for the $\chi_{c2}$.
The combined results from the three decay processes are also shown.
The first and second errors for the masses and widths are statistical 
and systematic, respectively. The other errors are statistical only.}
\ \\
\label{tab:table1}
\begin{tabular}{c|cccc} 
\hline
\hline
Charmonium $\to$ & Mass & Width($\Gamma$) (or $\sigma_M$)  & Number of events & $\chi^2/d.o.f$\\
 final state     &   (MeV/$c^2$)  & (MeV)  &   & \\    
\hline
$\eta_c \to 4\pi$ & $2987.5 \pm 1.7 \pm 2.0 $ & $36.0 \pm 5.0 \pm 1.7$ & $5381 \pm 492$ & 48.3/34\\
$\eta_c \to 2K2\pi$ & $2983.6 \pm 1.6 \pm 2.4$& $23.2 \pm 5.1 \pm 2.8$ & $2019 \pm 248$ & 61.7/34\\
$\eta_c \to 4K$ & $2988.0 \pm 2.3 \pm 4.0$ & $19.0 \pm 7.8 \pm 2.2$ & $216 \pm 42$ & 35.5/34\\
\hline
$\chi_{c0} \to 4\pi$ & $3414.7 \pm 0.7 \pm 2.0$ & $11.1 \pm 2.6 \pm 2.3$ & $ 3550 \pm 273$ & 27.1/14\\
$\chi_{c0} \to 2K2\pi$ & $3413.0 \pm 0.8 \pm 2.3$ & $10.1 \pm 3.0 \pm 3.1$ & $1694 \pm 161$ & 24.8/14\\
$\chi_{c0} \to 4K$ & $3415.6 \pm 2.2 \pm 3.1$ & $10.1 \pm 7.1 \pm 2.2$ & $215 \pm 36$ & 24.7/14\\
\hline
$\chi_{c2} \to 4\pi$ & $3554.4 \pm 0.9 \pm 2.0$ & ($9.2 \pm 0.8$) &  $1597 \pm 138$ & 4.7/4\\
$\chi_{c2} \to 2K2\pi$ & $3555.0 \pm 1.0 \pm 2.1$ & ($8.4 \pm 0.8$) & $780 \pm 74$ & 2.8/4\\
$\chi_{c2} \to 4K$ & $3558.0 \pm 1.8 \pm 2.9$ & ($8.1 \pm 1.3$) & $126 \pm 24$ & 3.9/4\\
\hline
$\eta_c$ (combined) & $2986.1 \pm 1.0 \pm 2.5$ & $28.1 \pm 3.2 \pm 2.2$ & \\
$\chi_{c0}$ (combined) & $3414.2 \pm 0.5 \pm 2.3$ & $10.6 \pm 1.9 \pm 2.6$ & \\
$\chi_{c2}$ (combined) & $3555.3 \pm 0.6 \pm 2.2$ & $-$ & \\
\hline
\hline
\end{tabular}
\end{table}

\subsection{Yields of two-body decays}
We study the sub-structure of the charmonium signals by searching for
quasi-two-body components, in which
a charmonium meson decays into two resonances, and each
resonance decays to two final-state mesons. 
 We determine the yields of these two-body decays 
using the following procedure: we first make  
two-dimensional plots of the two invariant masses constructed from the
available two-meson combinations of $\pi^+\pi^-$, $K^{\pm}\pi^{\mp}$ or
$K^+K^-$, for each event.  The $4\pi$ and $4K$ samples
have two entries from each event. Such plots are
made in each charmonium signal region and sideband region.
The charmonium sideband distribution is then subtracted
from the signal distribution bin-by-bin in two dimensions.
We note that the non-charmonium background components are considerable
even in the vicinities of the charmonium peaks.  

The signal regions
are defined as the ranges within $\pm 50$~MeV/$c^2$ ($\pm 30$~MeV/$c^2$ for 
the $\chi_{c2}$)
of 2.98~GeV/$c^2$, 3.41~GeV/$c^2$, and 
3.56~GeV/$c^2$, for the $\eta_c$, $\chi_{c0}$ and $\chi_{c2}$, respectively.
The sideband regions are taken on both sides of the
signal region, $[-100, -50]$~MeV/$c^2$ below and $[+50, +100]$~MeV/$c^2$
above ( $[-60, -30]$~MeV/$c^2$ and $[+30, +60]$~MeV/$c^2$
for the $\chi_{c2}$) the central point of the corresponding signal region.
The sizes of the signal and sideband regions
(in the sum of the two regions) in the invariant-mass range
are chosen to be the same so that we can apply a simple sideband subtraction. 

Next, we define the signal and sideband regions for the 
$\rho^0 \to \pi^+\pi^-$, $f_2(1270) \to \pi^+\pi^-$, 
$K^*(892)^0 \to K^{\pm}\pi^{\mp}$ and 
$\phi \to K^+K^-$ mesons, based on their known masses and 
widths. (Hereafter, we refer to the $f_2(1270)$, $K^*(892)^0$ and 
$f'_2(1525)$ as $f_2$, $K^{*0}$ and $f'_2$, respectively.)
We apply the sideband subtraction at the $f_2$ side to extract
the yield of the $f_2 f'_2$ decay.

 The signal regions are 0.64-0.88~GeV/$c^2$, 1.08-1.40~GeV/$c^2$,
0.80-0.96~GeV/$c^2$ and 1.00-1.04~GeV/$c^2$ for the $\rho^0$, $f_2$,
$K^{*0}$ and $\phi$, respectively.
We take sideband regions on both sides adjacent to the signal
region. In the $\phi \to K^+K^-$ mode the sidebands cover the ranges from
the mass threshold to 1.00~GeV/$c^2$ and 1.04-1.08~GeV/$c^2$.
For each of the other resonances the sideband has a 
width that is half the size of the signal region on each side. The sideband
subtraction is applied assuming the continuum yield has
a linear dependence on invariant mass.

Thus, we obtain a one-dimensional histogram
corresponding to a resonance signal component 
tagged by another resonance. We search for $\rho^0$, $f_2$, $K^{*0}$,
$\phi$ and $f'_2 \to K^+K^-$ signals in the histograms.
In the decays to pairs of same-kind resonances (such as
$f_2 f_2$ or $K^{*0}\bar{K}^{*0}$), we symmetrized the two-dimensional
distribution in two directions, by adding it 
to the same distribution where the two-masses
are interchanged. In this case we must divide the peak yield 
by two to obtain the correct signal yield and take into account
the statistical correlation.

We fit the signal to a Breit-Wigner taking into account
the angular momentum between the final state mesons
and a linear continuum component in the one-dimensional
distribution near the resonance masses.  
As the only exception, however, we count in  
$\phi\phi$ decays
the number of events in the $K^+K^-$ invariant
mass range, 1.00-1.04~GeV/$c^2$ and subtract backgrounds 
from a sideband in the 1.04-1.24~GeV/$c^2$ region,
instead of performing a fit.
The one-dimensional distributions and the fits for the $\rho^0$, $f_2$,
$K^{*0}$ and $f'_2$ resonances in different charmonium
decay modes are shown in Figs.~5 - 7. 

 Table~\ref{tab:table2} summarizes the signal yields.
We assume that there are no effects from interference in the 
sideband subtractions as well as the fits of the invariant
mass distributions.
We have applied corrections for
inefficiencies arising from sideband subtractions 
in the two-meson resonance 
and charmonium regions; the targeted resonance component partially 
leaks outside the signal region and enters the sideband regions. 
The inefficiencies are calculated
by assuming Breit-Wigner forms for the charmonia and
light-quark resonances using the known masses and widths~\cite{pdg}.
The 90\%-confidence-level (C.L.) upper limits correspond to 
the upper edge of a 90\% confidence region derived from Feldman and Cousins's
table (Table X in Ref.~\cite{felcou}) 
assuming statistical fluctuations follow Gaussian 
distributions (there is a large number of events
in both signal and sideband regions in the present case), 
with the expected number of signal events constrained to 
be non-negative.

We do not observe a significant $\rho^0\rho^0$ component
in $\eta_c$ decays in spite of the signal reported for
$\eta_c \to \rho\rho$ from measurements of 
radiative $J/\psi$ decays~\cite{pdg}.  
Meanwhile, we find  $f_2 f_2$ and $f_2 f'_2$ signals with 
greater than $4\sigma$ statistical significances. 
It can be noted that the observed yield for $\eta_c \to f_2 f_2 \to 4\pi$ 
is comparable to the total yield 
of $\eta_c \to 4\pi$.  This indicates that the
$f_2f_2$ component dominates  
$\eta_c \to 4\pi$ decays.  We confirm the decays to
$f_2 f_2$ in the distribution of the transversity angle (Fig.~8(a)) defined
as the angle between the decay planes of the two 
$f_2$ candidates in the $\eta_c$
rest frame. Here, the $f_2$ candidate is a $\pi^+\pi^-$ 
combination whose invariant mass lies in the range 1.15-1.39~GeV/$c^2$.  
The distribution shows a characteristic feature of tensor-meson
decays.
Similarly, $\eta_c \to 2K2\pi$ is dominated by the sum of
$\eta_c \to K^{*0}\bar{K}^{*0} \to 2K2\pi$ and
$\eta_c \to f_2 f'_2 \to 2K2\pi$. The transversity-angle
distribution of the $f_2 f'_2$ candidates is shown in Fig.~8(b)
(here, the $f'_2$ candidate is a $K^+K^-$ combination whose invariant mass falls
in 1.475 - 1.575~GeV/$c^2$).
The asymmetry with respect to 90$^\circ$
can be interpreted as the effect of interference between the above 
two decay modes; an asymmetry in directional
correlations between $K^+\pi^+$ and  $K^+\pi^-$ should 
not arise in a pure $f_2 f'_2$ process. 
The ($M(K^-\pi^+)$, $M(K^+\pi^-)$) distributions expected from the $f_2 f'_2$ 
MC events overlap with the ($K^{*0}$, $\bar{K}^{*0}$) mass region near 
its lowest mass-combination edge.
The histograms are the distributions
of the corresponding signal MC events generated assuming
pure P-waves between the two tensor mesons.

Figures~9 (a) and (b) show the invariant-mass distributions
of $4\pi$ events that are consistent with $\rho^0\rho^0$ and
$f_2 f_2$, respectively.
We find no prominent peaks at the $\eta_c$ mass in the former, but find
a clear signal for the $\eta_c$ in the latter. An enhancement
at the $\eta_c$ mass is also clearly visible in Fig.~9(c) which
shows the invariant-mass distribution for $f_2 f'_2 \to 2K2\pi$
candidates.  

\begin{table}[htb]
\caption{
The yields of two-body and three-body decay events.
The errors are statistical. The efficiency due to the 
sideband subtraction procedure is corrected.  The upper limits are
at the 90\%~C.L.\\}
\label{tab:table2}
\begin{tabular}{c|c} 
\hline
\hline
Decay process  & Number of events \\  
\hline
$\eta_c \to \rho^0\rho^0 \to 4\pi$ & $<1556$  \\
$\eta_c \to f_2 f_2 \to 4\pi$ &  $3182 \pm 766$ \\
$\chi_{c0} \to \rho^0\rho^0 \to 4\pi$ & $<252$ \\
$\chi_{c2} \to \rho^0\rho^0 \to 4\pi$ & $<598$  \\
$\eta_c \to K^{*0}\bar{K}^{*0} \to 2K2\pi$ &  $882 \pm 115$ \\
$\eta_c \to f_2 f_2' \to 2K2\pi$ &  $1128 \pm 206$  \\
$\chi_{c0} \to K^{*0}\bar{K}^{*0} \to 2K2\pi$ &  $<148$ \\
$\chi_{c2} \to K^{*0}\bar{K}^{*0} \to 2K2\pi$ &  $151 \pm 30$ \\
$\eta_c \to \phi\phi \to 4K$ &   $132 \pm 23 $\\
$\chi_{c0} \to \phi\phi \to 4K$ &  $23.6 \pm 9.6$ \\
$\chi_{c2} \to \phi\phi \to 4K$ &  $26.5 \pm 8.1$ \\
\hline
$\chi_{c2} \to \rho^0\pi^+\pi^- \to 4\pi$ &  $986 \pm 578$ \\
$\chi_{c0} \to K^{*0}K^-\pi^+ {\rm (or\ c.c.)} \to 2K2\pi$ &  $495 \pm 182$ \\
\hline
\hline
\end{tabular}
\end{table}

\begin{figure}[tb]
\begin{minipage}[t]{78mm}
\begin{center}
\includegraphics[width=8cm]{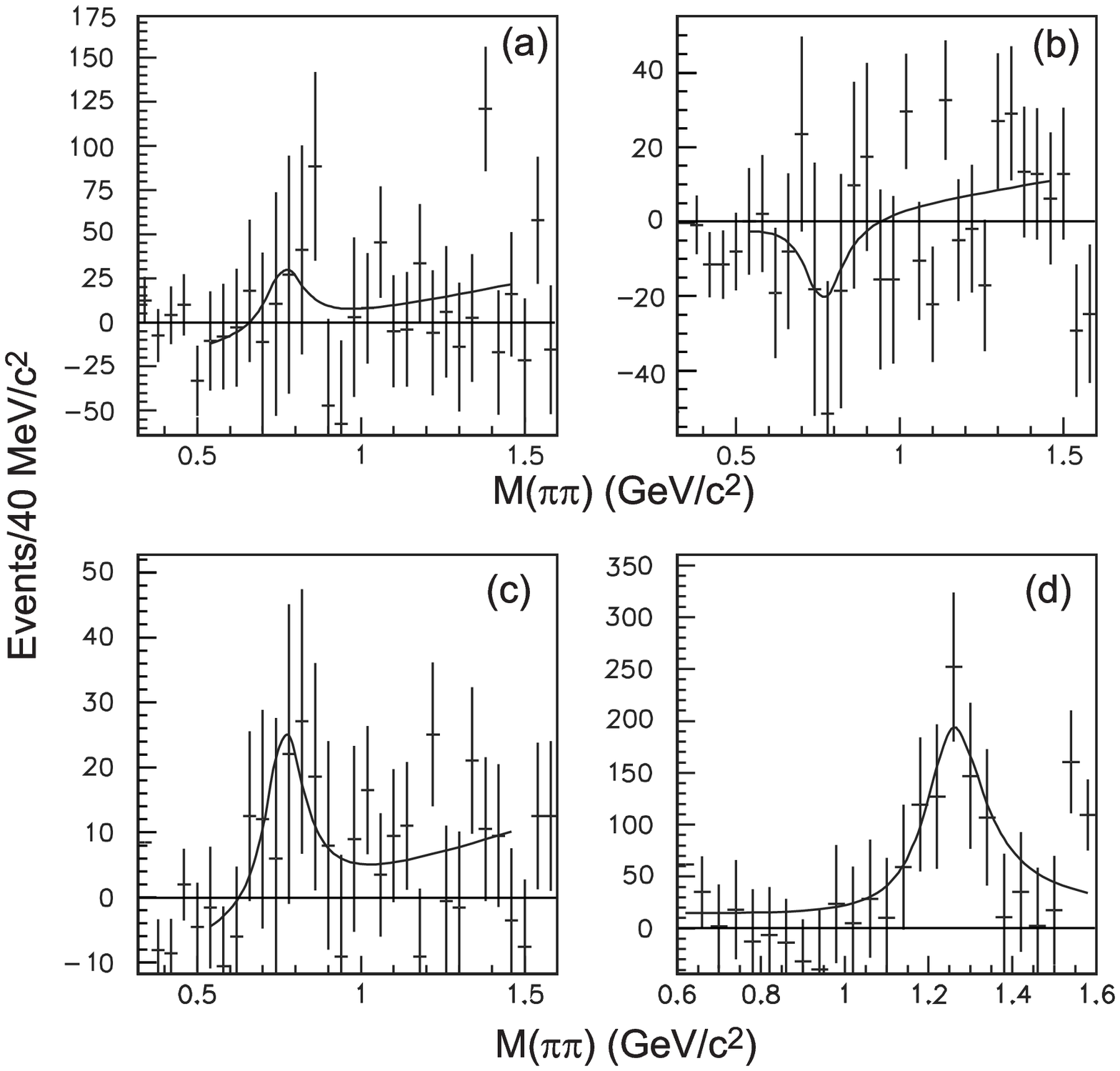}
\label{fig13}
\caption{The two-meson invariant-mass 
distributions and fits used to determine 
two-body yields in 
(a) $\eta_c \to \rho^0 \rho^0$, 
(b) $\chi_{c0} \to \rho^0 \rho^0$, 
(c) $\chi_{c2} \to \rho^0 \rho^0$, and
(d) $\eta_c \to f_2 f_2$. A $\rho^0$ or a $f_2$ is selected
on the opposite side. We rescale the number of entries described in 
Sect.~III.B by a factor of 1/2 to normalize 
the vertical axis to the number of events. 
}
\end{center}
\end{minipage}
\hspace{\fill}
\begin{minipage}[t]{78mm}
\begin{center}
\includegraphics[width=8cm]{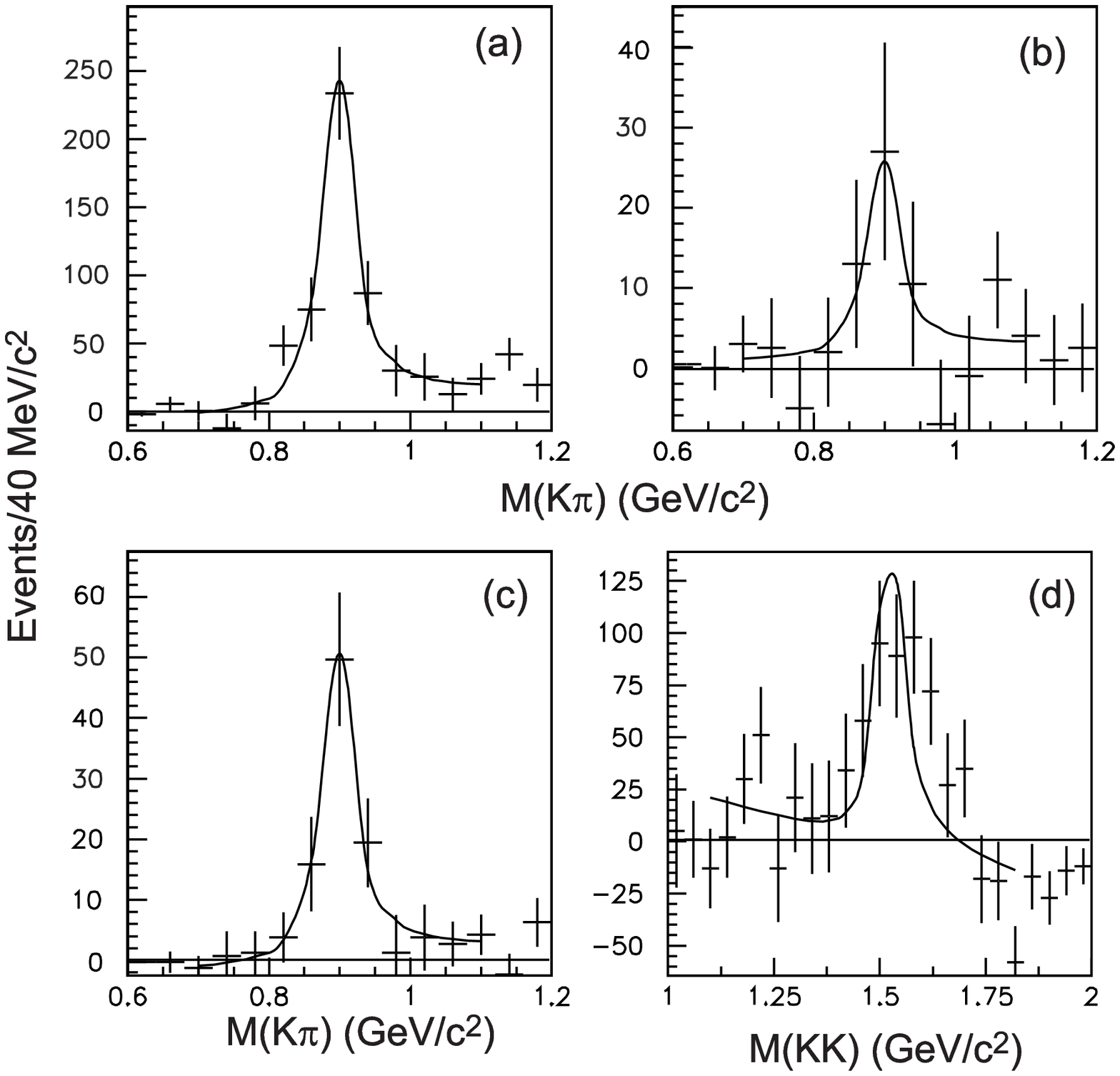}
\label{fig14}
\caption{The two-meson invariant-mass 
distributions and fits used to determine
two-body yields in (a) $\eta_c \to K^{*0}\bar{K}^{*0}$, 
(b) $\chi_{c0} \to K^{*0}\bar{K}^{*0}$, 
(c) $\chi_{c2} \to K^{*0}\bar{K}^{*0}$, and
(d) $\eta_c \to f_2 f_2'$. 
A $\bar{K}^{*0}$ or a $f_2$ is selected
on the opposite side. We rescale the number of entries described in 
Sect.~III.B by a factor of 1/2 to normalize 
the vertical axis to the number of events, except for (d). }
\end{center}
\end{minipage}
\end{figure}

\begin{figure}
\centering
\includegraphics[width=14cm]{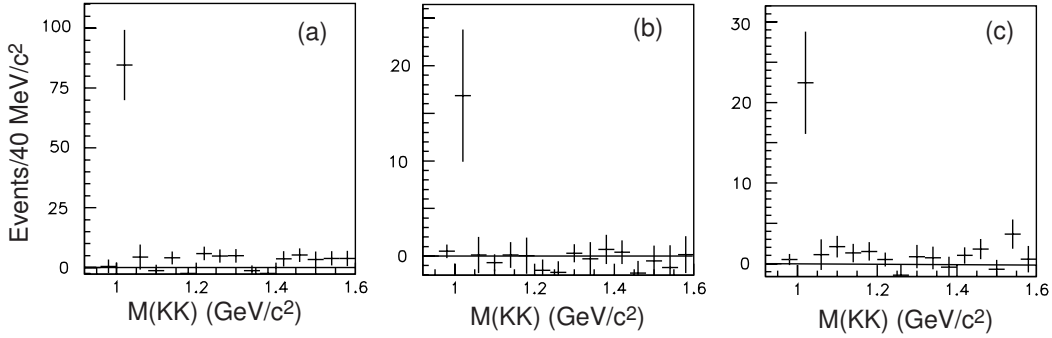}
\label{fig15}
\centering
\caption{The two-meson invariant-mass 
distributions in (a) $\eta_c \to \phi\phi$, 
(b) $\chi_{c0} \to \phi\phi$, and
(c) $\chi_{c2} \to \phi\phi$. A $\phi$ is selected
on the opposite side. We rescale the number of entries described in 
Sect.~III.B by a factor of 1/2 to normalize 
the vertical axis to the number of events.}
\end{figure}

\begin{figure}
\centering
\includegraphics[width=10cm]{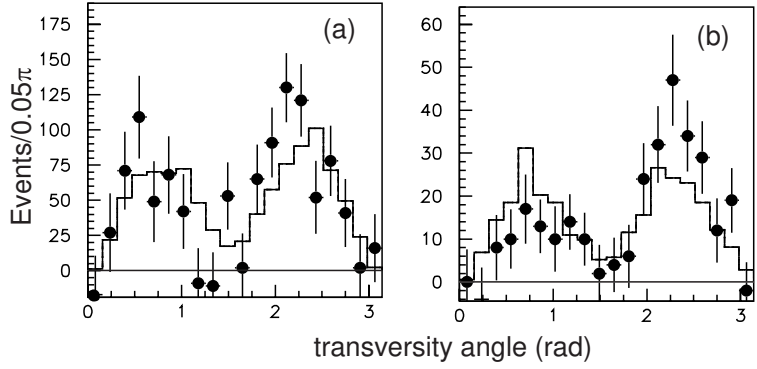}
\label{fig15}
\centering
\caption{The transversity-angle distributions for
(a) $\eta_c \to f_2 f_2 \to 4\pi$ and (b) $\eta_c \to f_2 f'_2 \to 2K2\pi$
candidates for the experimental data after the subtraction
of charmonium sideband contributions (closed circles with error bars). 
The histograms are the distributions from signal MC (see the text
in Sect.III.B). }
\end{figure}

\begin{figure}
\centering
\includegraphics[width=14cm]{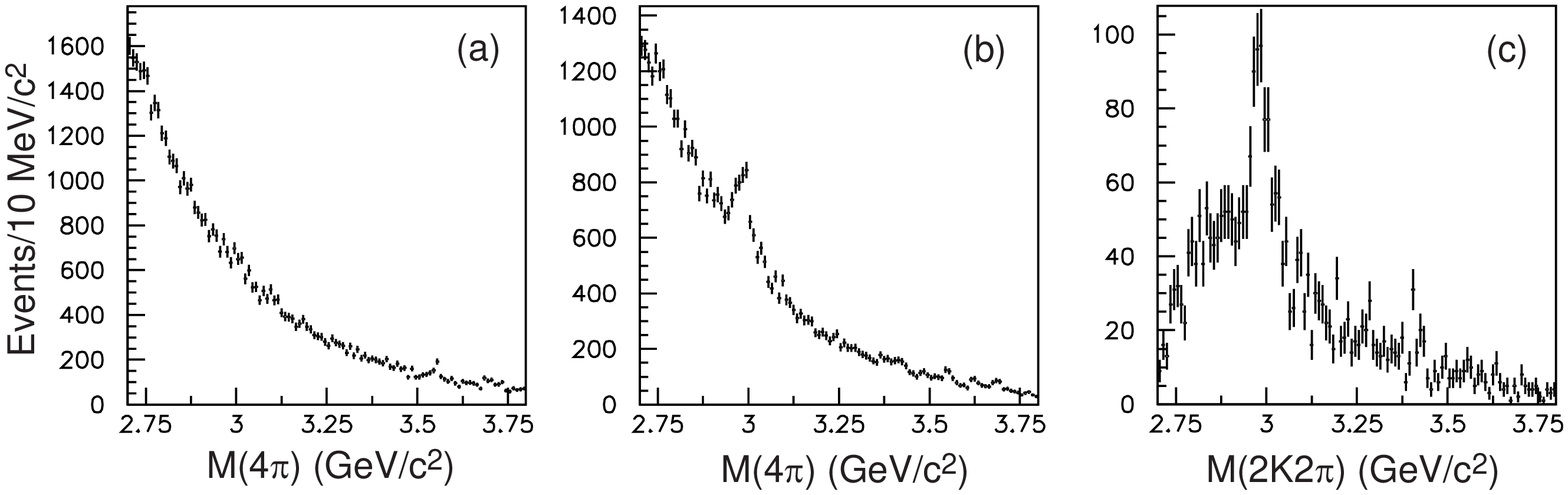}
\label{fig15}
\centering
\caption{The $4\pi$ invariant-mass distributions for 
(a) $\rho^0\rho^0$ and (b) $f_2 f_2$ candidates 
and the $2K2\pi$ invariant-mass distributions for 
(c) $f_2f'_2$ candidates.
No subtraction is applied for $\rho^0$, $f_2$ and  $f'_2$ 
sidebands, that is, contributions from $\rho^0 \pi^+\pi^-$ etc.
may also be included.}
\end{figure}

\subsection{Yields of three-body decays}
 The distributions corresponding to the processes 
 $\chi_{c2} \to \rho^{0}\pi^+\pi^-$  in Fig.~5(c)
and $\chi_{c0} \to K^{*0} K^-\pi^+$ or $\bar{K}^{*0}K^+\pi^-$ in Fig.~6(b)
have significant net yields although there are no
prominent structures from the two-body decay components,
$\chi_{c2} \to \rho^0\rho^0$ and $\chi_{c0} \to 
K^{*0}\bar{K}^{*0}$. We obtain 
the yields of each three-body decay using the following
procedure. For the two processes, we first obtain
the numbers of $\rho^0$'s in $\chi_{c2} \to 4\pi$
and  $K^{*0}$'s in $\chi_{c0} \to 2K2\pi$
by fitting the invariant-mass distributions of $\pi^+\pi^-$ 
(Fig.~10(a), 4 entries/event) and
$K^{\pm}\pi^{\mp}$ (Fig.~10(b),
2 entries/event) to functional forms corresponding to 
the sum of $\rho^0$ and $f_2$ resonances
and a $K^{*0}$ resonance, respectively. Here, we also take into account
the continuum component parameterized by a second-order polynomial. 
The results obtained are converted into the yields of
three-body decay events after subtracting possible resonance 
yields from $\chi_{c2} \to \rho^{0}\pi^+\pi^-$ via $\rho^0\rho^0$   
and  $\chi_{c0} \to K^{*0}K^+\pi^- $ (or c.c.) 
via $K^{*0}\bar{K}^{*0}$  using the best estimates of 
the corresponding two-body yields after doubling the number 
of events to take account of the multiple entries. 
The results after these subtractions are
summarized in Table~\ref{tab:table2}.

\begin{figure}
\centering
\includegraphics[width=10cm]{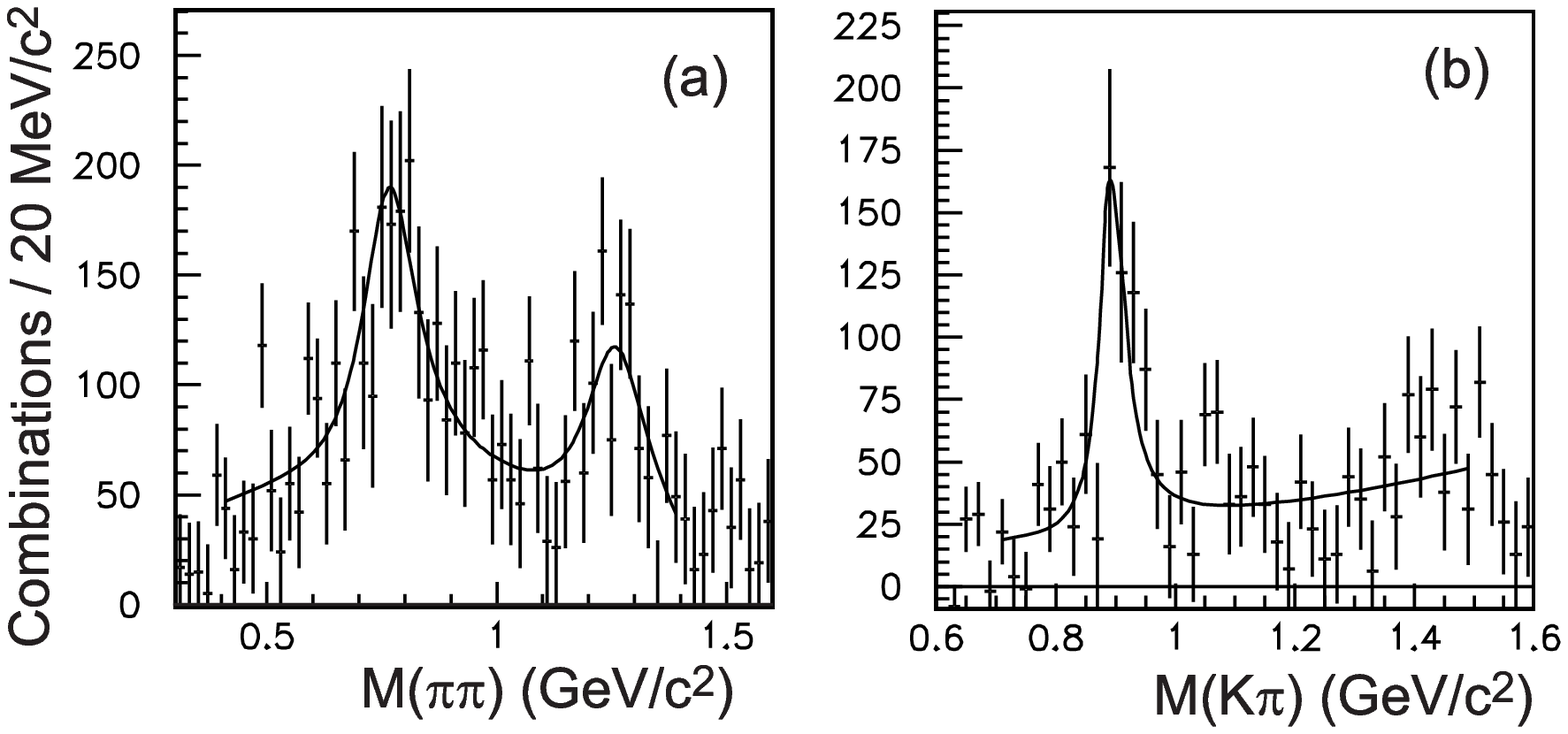}
\label{fig15}
\centering
\caption{(a) $\pi^+\pi^-$ invariant distributions in
the $\chi_{c2}$ decay products and (b) $K^{\pm}\pi^{\mp}$ 
invariant distributions in the $\chi_{c0}$ decay products.
The charmonium-sideband distributions are subtracted. The curves
are the fits considering the resonances decaying into
two mesons (see text in Sect.~III.C).}
\end{figure}

\section{Derivation of Products of Two-photon Decay Width and
Branching Fraction}

  The product of the two-photon decay width and branching fraction 
is obtained from the following formula:
\[
{\cal G}(R \to {\rm final~state}) \equiv \Gamma_{\gamma\gamma}(R){\cal B}
(R \to {\rm final~state}) = \frac{N}{
\varepsilon\ \int{\cal L}dt\ F(M_R, J)},
\]
where $N$ is the number of observed events, $\varepsilon$ the efficiency,
and $\int{\cal L}dt$ the integrated luminosity.  $F(M_R,J)$ is
a factor that is calculated from the two-photon luminosity 
function ($L_{\gamma\gamma}(M_R)$),
the charmonium mass ($M_R$) and the charmonium spin ($J$), using
the relation 
\[
F(M_R,J) = 4\pi^2(2J+1)L_{\gamma\gamma}(M_R)/{M_R}^2,
\]
that is valid when the resonance width is small compared
to the available kinetic energy in the decay.
The $F$ parameter is calculated to be 
2.11~fb/eV, 1.15~fb/eV, 4.74~fb/eV and 0.861~fb/eV for the 
$\eta_c$, $\chi_{c0}$, 
$\chi_{c2}$ and $\eta_c(2S)$, respectively, using the TREPS code~\cite{treps}. 
We take the signal yields from
Tables~\ref{tab:table1} and \ref{tab:table2}.  
We subtract the small contributions from $K^0_SK^0_S$ decays 
included in the $\chi_{c0} \to 4\pi$ and 
$\chi_{c2} \to 4\pi$ samples, $142\pm6$ and $35\pm5$ events respectively, 
which were determined from the 
two-dimensional invariant-mass plots mentioned in Sect.~III.A.
($\chi_{c0}$, $\chi_{c2} \to K^0_S K^0_S$ results from our 
experiment are reported separately~\cite{wtchen}.
$\eta_c \to K^0_S K^0_S$ is prohibited from parity
conservation.)

The efficiency in the detection of each process is determined
from the signal MC events generated
by the TREPS code~\cite{treps} and processed by a full detector simulation.
Three- and four-body decays were generated according to phase space with a
resonance mass distributed according to the Breit-Wigner function,
and the $K^{*0}$($\rho^0$) decaying to $K^{\pm}\pi^{\mp}$
($\pi^+\pi^-$) isotropically.

For two-body decays involving resonances with spin, the decay
amplitude takes into account correlations caused by spin, and the
full matrix element is symmetrized with respect to identical
final-state particles. 
We assume a pure state with the lowest possible orbital 
angular momentum that satisfies conservation laws
in each decay channel. The helicity 
of the $\chi_{c2}$ along the incident axis is assumed to be 2, following 
the previous experiment results~\cite{bellechic2} and theoretical 
expectations~\cite{chic2theo}.

The trigger efficiency is taken into account; it is estimated to be around
95\% within the acceptance for all the processes. The detection 
efficiency including that of particle identification
is calculated using 
MC events passed through the detector simulation code. 
Typical efficiencies are 16\%, 10\%, 
5\%, and 9\% in the $4\pi$ final-state processes, 
$2K2\pi$ final-state processes, $4K$ with a phase-space
distribution, and the $\phi\phi$ process, respectively.
Their dependence on the two-photon c.m. energy, or charmonium mass, 
is rather small.

We summarize the results on
${\cal G} \equiv \Gamma_{\gamma\gamma}{\cal B}$ in Table~\ref{tab:table3}.
The systematic errors and comparison
to the previous experimental results also given there
are described in detail in the following two sections.
In some decay channels, 
we assume isospin invariance and use necessary branching 
fractions from Ref.~\cite{pdg}.
The ``isospin $\cdot BF$'' factors in the table are the products
of the isospin factor and the branching fraction(s) used to obtain
the results.

\section{Systematic Errors} 
\subsection{Uncertainties in efficiencies}
  The trigger efficiency is estimated
using the Monte-Carlo trigger simulator to be $(95 \pm 4)\%$,
where the error is estimated from comparison
of experimental and MC results in low-multiplicity
events and the efficiency variation under different beam 
background conditions. 
 
We obtain a 5.5\% systematic error from the
uncertainty in track reconstruction efficiency
for the four-track events, as well as 2\% per 
kaon track from the kaon identification
efficiency. The latter is determined from a study of 
kinematically identified kaons in a sample of 
$D^{*+} \to D^0\pi^+$, $D^0 \to K^- \pi^+$ decays.  
We check the kaon identification efficiency by loosening
the $4K$-event selection requirement; we temporarily 
require only three kaons in an event. We obtain $(65 \pm 16)$\% 
more yield for the $\eta_c$, $\chi_{c0}$ and  $\chi_{c2}$,
on average with the looser condition.
The charmonium peaks are less  prominent with these selections,
due to a larger contamination from non-$4K$ events.
The above ratio is consistent with expectations from the MC, 
$\sim 53$\%, for the average of the three charmonia decaying 
to the $4K$ state.  

We use the total widths of the charmonia for
the determination of the efficiency in the sideband subtractions.
Taking systematic errors in 
the $\eta_c$ and $\chi_{c0}$ widths into account,  
we varied the widths of the $\eta_c$ and $\chi_{c0}$
from their nominal values by $\pm 10$~MeV and $\pm 7$~MeV, respectively, 
and assigned the obtained efficiency changes,
$\pm 12\%$ and $\pm 8\%$, respectively, as a systematic error.
We take a 3\% systematic error in the $\chi_{c2}$ case 
for the uncertainty in the treatment of its long Breit-Wigner tails. 

Inefficiencies caused by the sideband subtractions of the isobar resonances
decaying to two mesons are calculated using
the resonance parameters. We estimated the systematic error
from this effect by changing the integration range of the resonance
function from $\pm 8\Gamma$ to $\pm 5\Gamma$. 
It is estimated to be 5\% for each two-body decay mode.


\subsection{Systematic errors in signal yields}
In the fits to the four-meson invariant-mass spectra described in Sect.~III.A,
we set the fit region for the analyses of the $\eta_c$ to be rather wide, 
considering the relatively large natural width and the relatively large size
of the continuum components. As a result, we have  
a larger second-order polynomial term 
contribution from the continuum components, 
which correlates significantly with the size of the
charmonium component and could bias the signal yield. 
We tentatively narrowed the fit region to 2.85 - 3.15~GeV/$c^2$ and
took the variations in the obtained yields as a systematic error.
They amount to 7\%, 13\% and 24\% for $4\pi$, $2K2\pi$ and $4K$ channels,
respectively.  We neglect this error source for 
the other charmonia, where
the quadratic components in the continuum are small. 

In the fits to the two-meson invariant-mass distributions
used to obtain the yields of the two-body decays,  we 
tried a second-order
polynomial constrained to vanish at the two-body threshold
for each continuum component. We assign the difference between
the yields from this method and the standard fit 
(to a linear function) to the systematic error.
Thus, the systematic errors are 20\% for $\eta_c \to f_2 f'_2$ 
and less than 3\%
for the other two-body channels where we have significant
signals. 

In the case of the three-body modes, we tested an alternative
functional form for the continuum component and used 
the shape of three-body phase space calculated by the MC generator
and the detector simulator. We take the difference
in the signal yields from this method as the systematic error,
that is 4\% and 7\% for $\chi_{c2} \to \rho^0 \pi^+\pi^-$ and
$\chi_{c0} \to K^{*0}K^{\mp}\pi^{\pm}$, respectively.

\subsection{Study of non-exclusive backgrounds}
Backgrounds can come from general multi-hadron
production in two-photon collisions 
and $q\bar{q}$ production in single-photon 
annihilation.  
Yields from the $c\bar{c}$ process in the latter 
category are estimated from the MC and 
found to be negligibly small ($\sim 1\%$) 
compared to the non-charmonium backgrounds from
two-photon processes. 
However, the charmonium production rates from these processes
cannot be reliably estimated.

  A more reliable signature is the $p_t$-balance nature in the signal events.
We expect that non-$2\gamma$ 
background processes cannot produce a $p_t$-balanced 
distribution, except for double-ISR processes
where exclusive production of $C=+$ charmonia 
is prohibited.
Even in the decays $\psi(2S) \to \gamma\chi_{cJ}$ where
a $\psi(2S)$ is produced in an ISR or double ISR process, 
the $p_t$ distribution of $\chi_{cJ}$ will not  
peak inside the selected region, 
$|\Sigma {\bf p}^*_t| < 0.1$~GeV/$c$. We estimate the yield of
this process to be about 1\% of the $\chi_{c0}$ and $\chi_{c2}$
signals, based on the observed yields of $\psi(2S)$ 
decaying to $\pi^+\pi^- J/\psi$.

We evaluate the non-exclusive backgrounds
by examining the observed $|\Sigma {\bf p}^*_t|$ distributions.
Figure~11 shows the $|\Sigma {\bf p}^*_t|$ distributions
of the charmonium signal component after the
sideband subtractions. The distributions are
compared with the signal-MC expectations normalized to
the sum of the two leftmost bins where
the background contamination is expected to be
very small.  We expect that the background component has a 
linear shape in $|\Sigma {\bf p}^*_t|$, in the region below 0.1~GeV/$c$,
based on geometrical considerations and some MC studies.
Given this assumption, we obtain the background level
from data, to be $(6 \pm 7)$\%, $(-4 \pm 5)$\% and $(9 \pm 7)$\%, 
on average for the $\eta_c$, $\chi_{c0}$ and $\chi_{c2}$
decay modes, respectively, using the experimental excesses
seen in the bins of $0.07 < |\Sigma {\bf p}^*_t| < 0.10$~GeV/$c$. 
Since there is no significant difference
among these estimates, and they are consistent with zero 
(the overall average is $(2 \pm 4)$\%), we do not 
apply any corrections to this background source
and assign a systematic error of 6\% for all processes. 

\begin{figure}
\centering
\includegraphics[width=10cm]{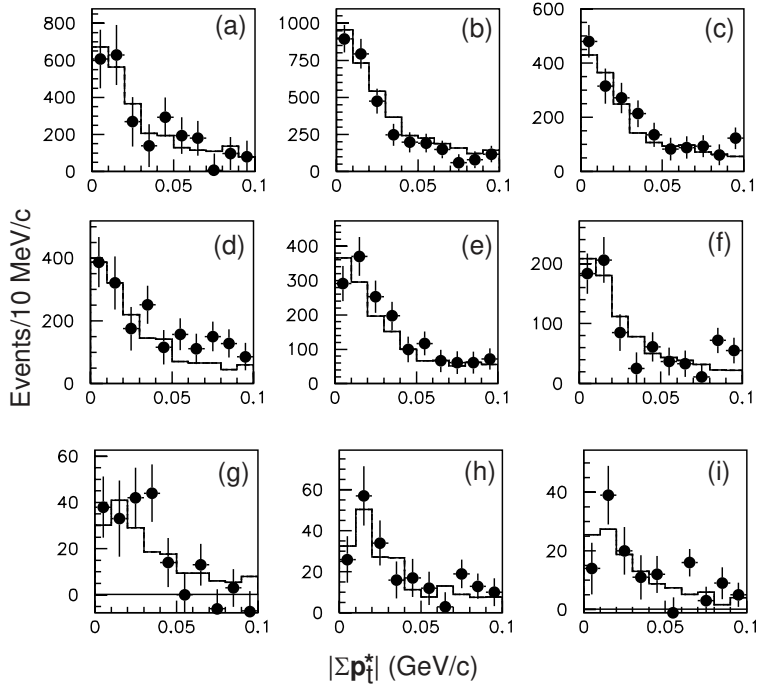}
\label{fig21}
\centering
\caption{The $|\sum {\bf p}_t^*|$ distributions for the charmonium
signal components after the charmonium-sideband subtractions (closed circles
with error bars), 
compared with the signal MC (histograms) in the processes 
(a)$\eta_c \to 4\pi$,  (b)$\chi_{c0} \to 4\pi$, 
(c)$\chi_{c2} \to 4\pi$,  (d)$\eta_c \to 2K2\pi$,  (e)$\chi_{c0} \to 2K2\pi$, 
(f)$\chi_{c2} \to 2K2\pi$,  (g)$\eta_c \to 4K$,  (h)$\chi_{c0} \to 4K$ and
(i)$\chi_{c2} \to 4K$.
The MC is normalized to the sum of the two leftmost bins.}
\end{figure}

\subsection{Total systematic errors}
We consider other sources of systematic errors. The statistics
of the signal MC events gives errors of 2-4\% depending on
the process. The calculation of the luminosity function
has a systematic uncertainty of 5\%. We take the differences 
among the efficiency estimates for
different spin/angular-momentum assumptions for the 
$\chi_{c2}$ decays to $K^{*0} \bar{K}^{*0}$ and
$\phi\phi$ as systematic errors. The nominal S-wave hypothesis
gives the most conservative upper limit for the  
$\chi_{c2} \to \rho^0 \rho^0$ case.
The total systematic errors are  listed in Table~III.  
The size of the systematic error relative to the central value is process
dependent and ranges from 11\% to 35\%
in the processes where finite ${\cal G}$ results
are presented.  
The upper limits in the table are shifted upwards 
by the $1\sigma$ systematic errors, in order to include the effect
of systematic uncertainties.

\section{Comparison with Previous Measurements}

 Some previous measurements give ${\cal G}$ results 
for the processes measured here~\cite{delphi,cleo,argus}. 
We make direct comparisons with
these measurements, which are listed in
Table~\ref{tab:table3}. We quote the world average value from 
Ref.~\cite{pdg}
when two or more previous measurements are available. We can
make such direct comparison only for a limited number of processes,
since our results include many processes measured
for the first time.

It is possible to make some indirect comparisons  
by converting the ${\cal G}$ results measured for
a different decay mode from the present measured mode and multiplying 
by the ratio of the branching fractions;
\[
{\cal G}(R \to X) = {{\cal B}(R \to X)}\frac{{\cal G}(R \to A)}{{\cal B}(R \to A)},
\]
where $R \to A$ is a ``normalization'' process for which previous 
measurement(s) are available.  We adopt $\eta_c \to K \bar{K} \pi$,
$\chi_{c0} \to \pi^+\pi^-/K^+K^-$ (the average of the 
two modes) and $\chi_{c2} \to \gamma J/\psi$
as the normalization processes.  Note that the two-photon decay
widths for these charmonium states are so far measured only in a few 
decay modes; there are not enough consistency checks among the 
different decay modes available so far.
We use the PDG's fit when the average is unavailable~\cite{pdg},
and treat the errors shown there as independent from each other.

We find large differences
between the present and previous measurements for $\eta_c$
in the direct comparisons, although they
are not inconsistent considering the large errors
assigned to the previous measurements. 
However, in the indirect measurements, for many $\eta_c$ decay modes 
our ${\cal G}$ values are consistently smaller by factors of two to four compared to
the previous determinations. We do not observe any significant
signal in the $\rho^0\rho^0$ mode in contrast with previous
results~\cite{dm2,bes2}.
We conclude that these discrepancies are 
not likely to be due to statistical fluctuations, 
and there could be systematic deviations in
previous measurements of the relevant partial decay 
widths or branching fractions.

In contrast, all the present $\chi_{c0}$ and $\chi_{c2}$ 
results are in agreement with previous values, and 
confirm the values of $\Gamma_{\gamma\gamma}(\chi_{c0})$ 
and $\Gamma_{\gamma\gamma}(\chi_{c2})$
results derived in previous experiments~\cite{pdg}.  

\begin{table}[htb]
\caption{
The results in this paper for 
${\cal G} \equiv \Gamma_{\gamma\gamma}{\cal B}$ and
comparisons with previous measurements (shown as ``direct'' and 
``indirect'' below, see text)~\cite{cleo, argus, pdg}. The results with inequalities
correspond to 90\%-C.L. upper limits, where we include the systematic
errors by shifting the upper limits upwards by $1\sigma(sys)$.
The ``isospin $\cdot BF$'' factors are used to obtain the 
${\cal G}$ results from the experimental measurements.}
\ \\
\label{tab:table3}
\begin{tabular}{c|cc|cc} 
\hline
\hline
Process & Isospin $\cdot BF$ & This paper & Direct & Indirect   \\ 
      & factor & ${\cal G}$ (eV) & ${\cal G}$ (eV) & ${\cal G}$ (eV)\\   
\hline
$\eta_c \to \pi^+\pi^-\pi^+\pi^-$ & 1.0 & $40.7 \pm 3.7 \pm 5.3$ & $180 \pm 70 \pm 20$ & $83 \pm 24$\\
$\eta_c \to K^+K^-\pi^+\pi^-$ & 1.0 & $25.7 \pm 3.2 \pm 4.9$ & $210 \pm 70$ & $102 \pm 30$ \\
$\eta_c \to K^+K^-K^+K^-$ & 1.0 & $5.6 \pm 1.1 \pm 1.6$ & $280 \pm 70$ & $11 \pm 5$ \\
$\eta_c \to \rho\rho$ & 0.333 & $<39$ & $-$ & $130 \pm 43$ \\
$\eta_c \to f_2f_2$ & 0.320 & $69 \pm 17 \pm 12$ & $-$ & $74 \pm 36$ \\
$\eta_c \to K^*\bar{K}^*$ & 0.333 & $32.4 \pm 4.2 \pm 5.8$ & $-$ & $66 \pm 22$ \\
$\eta_c \to f_2f_2'$ & 0.251 & $49 \pm 9 \pm 13$ & $-$ & $-$ \\
$\eta_c \to \phi\phi$ & 0.243 & $6.8 \pm 1.2 \pm 1.3$ & $-$ & $19 \pm 5$ \\
\hline
$\chi_{c0} \to \pi^+\pi^-\pi^+\pi^-$ & 1.0 & $44.7 \pm 3.6 \pm 4.9$ & $75 \pm 13 \pm 8$ & $69 \pm 13$  \\
$\chi_{c0} \to K^+K^-\pi^+\pi^-$ & 1.0 & $38.8 \pm 3.7 \pm 4.7$ & $-$ & $53 \pm 12$ \\
$\chi_{c0} \to K^+K^-K^+K^-$ & 1.0 &  $7.9 \pm 1.3 \pm 1.1$ & $-$ & $7.8 \pm 1.6$\\
$\chi_{c0} \to K^{*0}K^-\pi^+ {\rm or \ c.c.}$ & 0.667 &  $16.7 \pm 6.1 \pm 3.0$  & $-$ & $34 \pm 13$    \\
$\chi_{c0} \to \rho\rho$ & 0.333 & $<12$ &  $-$ & $-$\\
$\chi_{c0} \to K^*\bar{K}^*$ & 0.333 & $<18$ & $-$ & $5.1 \pm 1.9$\\
$\chi_{c0} \to \phi\phi$ & 0.243 & $2.3 \pm 0.9 \pm 0.4$ & $-$ & $2.7 \pm 0.8$\\
\hline
$\chi_{c2} \to \pi^+\pi^-\pi^+\pi^-$ & 1.0 & $5.01 \pm 0.44 \pm 0.55$ & $6.4 \pm 1.8 \pm 0.8$ & $7.2 \pm 1.2$ \\
$\chi_{c2} \to K^+K^-\pi^+\pi^-$ & 1.0 & $4.42 \pm 0.42 \pm 0.53$ & $-$ & $5.8 \pm 2.1$\\ 
$\chi_{c2} \to K^+K^-K^+K^-$ & 1.0 & $1.10 \pm 0.21 \pm 0.15$ & $-$ & $1.03 \pm 0.18$\\ 
$\chi_{c2} \to \rho^0\pi^+\pi^-$ & 1.0 & $3.2 \pm 1.9 \pm 0.5$ & $-$ &  $3.9 \pm 2.3$\\
$\chi_{c2} \to \rho\rho$ & 0.333 & $<7.8$ &  $-$ & $-$\\
$\chi_{c2} \to K^*\bar{K}^*$ & 0.333 & $2.4 \pm 0.5 \pm 0.8$ & $-$ & $2.2 \pm 0.5$\\
$\chi_{c2} \to \phi\phi$ & 0.243 & $0.58 \pm 0.18 \pm 0.16$ &  $-$ & $1.0 \pm 0.3$\\
\hline
$\eta_c(2S) \to \pi^+\pi^-\pi^+\pi^-$ & 1.0 & $<6.5$ & $-$ & $-$ \\
$\eta_c(2S) \to K^+K^-\pi^+\pi^-$ & 1.0 & $<5.0$   &  $-$ & $-$ \\
$\eta_c(2S) \to K^+K^-K^+K^-$ & 1.0 & $<2.9$ & $-$ & $-$ \\
\hline
\hline
\end{tabular}
\end{table}

\section{Upper limits for $\eta_c(2S)$ Production}
  We search for the $\eta_c(2S)$ in the invariant-mass
distributions of the four-meson final states. Since the 
$\pi^+\pi^-\pi^+\pi^-$ final-state
sample has a prominent $\psi(2S)$ peak just above the $\eta_c(2S)$ mass
region, we veto events when the  
invariant mass of any $\pi^+\pi^-$ combination 
falls within 0.1~GeV/$c^2$ of
the nominal $J/\psi$ mass. (The two tracks from the $J/\psi$
decay are misidentified leptons.)  

We obtain upper limits on 
$\eta_c(2S)$ yields from the four-meson invariant-mass 
distributions (Fig.~12).   The mass and
width of the $\eta_c(2S)$ are not very precisely determined so far.
We consider a wide range for the mass and width, 
3.62~GeV/$c^2 < M(\eta_c(2S)) < 3.67$~GeV/$c^2$ and
10~MeV$<\Gamma(\eta_c(2S))<40$~MeV, according to previous 
measurements~\cite{belleetac2s,belledccbar,cleo2s,babar,pdg}.
Fits similar to those 
made in Sect.~III.A are applied for the $\eta_c(2S)$ 
using a Breit-Wigner function with
the mass resolution fixed to 9~MeV and
a second-order polynomial for the background component, as well as a Gaussian
function for the  $\chi_{c2}$  peak.  The upper limits for the yields are 
340, 164 and 55 events at 90\% C.L., for the $4\pi$, $2K2\pi$ and
$4K$ final-states, respectively.  The curves corresponding to
these upper limits are shown by the solid curves in Fig.~12. 
We calculate the upper limits using $\chi^2$ values from fits.
We assume various non-negative values of the $\eta_c(2S)$ yield, and,
for each case, we obtain the $\chi^2$ from the best fit while
floating all the other parameters.  We define the 90\% C.L. upper 
limit of the yield as one whose $\chi^2$ is larger by $(1.64)^2$
than the minimum $\chi^2$ derived in the different assumptions
of the yield. An application of the Feldman-Cousins method 
(Table X in Ref.~\cite{felcou}) to
the yields obtained from the fits gives the values 
very close to the yield upper limits, 
338, 163 and 55 events for the $4\pi$, $2K2\pi$ and $4K$ modes, 
respectively.

We calculate the efficiency assuming a phase space distribution.
The upper limit for ${\cal G} \equiv \Gamma_{\gamma\gamma}{\cal B}$ 
for each decay mode of the $\eta_c(2S)$ 
is shown in Table~\ref{tab:table3}.
 We find that the ratios  ${\cal G}(\eta_c(2S) \to {\rm four\ mesons})/
{\cal G}(\eta_c(1S) \to {\rm four\ mesons})$
are much smaller than unity in all three processes,
as previously found in another decay mode, $\eta_c(1S,~2S) \to
K^0_S K^\mp \pi^\pm$~\cite{cleo2s}.  
Although these results are still marginally   
consistent with the theoretical expectations,
$\Gamma_{\gamma\gamma}(\eta_c(2S))/
\Gamma_{\gamma\gamma}(\eta_c(1S)) \sim 0.3 $ and 
${\cal B}(\eta_c(2S) \to {\rm some\ hadronic\ final\ state})/
{\cal B}(\eta_c(1S) \to {\rm the\ same\ final\ state}) \sim 1$,
they give significant constraints on theoretical models~[1-3],
including calculations based on relativistic $q\bar{q}$
production.

\ \\
\begin{figure}
\centering
\includegraphics[width=10cm]{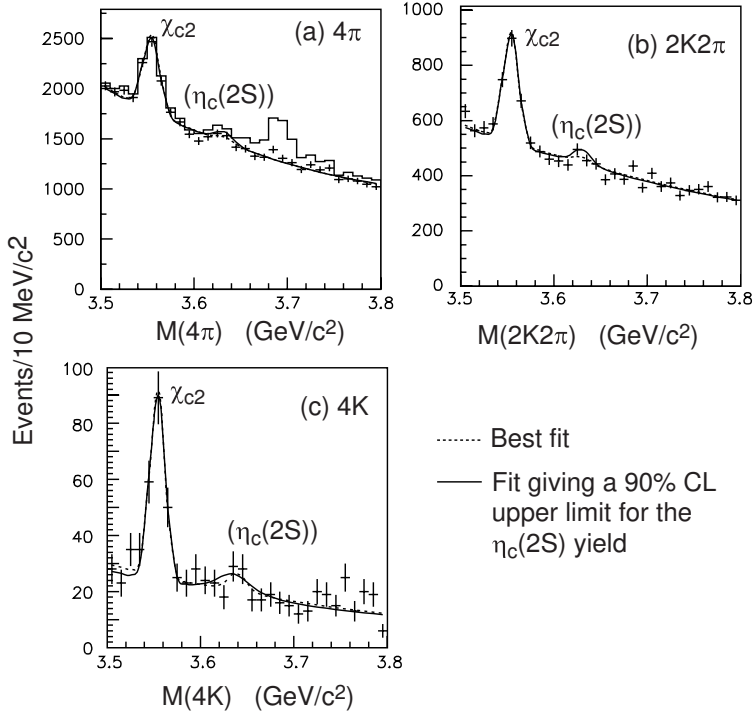}
\label{fig21b}
\centering
\caption{The invariant-mass distribution in the vicinities of 
$\eta_c(2S)$ mass 
in the final states, (a)$4\pi$, (b)$2K2\pi$ and (c)$4K$.  
The histogram and plots in (a) are the distributions 
before and after the rejection of $\psi(2S) \to J/\psi \pi^+\pi^-$, 
respectively. The dotted and solid curves show the best fit to the experimental
data and the fit with the $\eta_c(2S)$ yield set to the obtained
90\%-C.L. upper limits, respectively.}
\end{figure}

\section{Conclusion}

The production of the 
$\eta_c(1S)$, $\chi_{c0}$ and  $\chi_{c2}$
charmonium states in two-photon collisions 
has been observed in all of the four-meson
final states, $\pi^+\pi^-\pi^+\pi^-$, $K^+K^-\pi^+\pi^-$ and
$K^+K^-K^+K^-$.  We used data samples with 
one or two orders of magnitude larger statistics than previous measurements.
No clear signature for
the $\eta_c(2S)$ is found in any of decay processes, and
we obtain the upper limits for the products of its
two-photon decay width and the branching fractions.
We have studied resonant substructures in these four-meson 
final states.

For the first time $\chi_{cJ}$ signals produced
in two-photon collisions are observed in the $K^+K^-\pi^+\pi^-$
or $K^+K^-K^+K^-$ final states.  We also find a new decay
mode, $\eta_c \to f_2(1270)f'_2(1525)$.
We have obtained products of the two-photon decay width
and branching fractions for various decays 
of charmonium states. The present results for $\eta_c$ are systematically 
smaller than the derived values from the world averages of 
previous measurements.\\

\ \\
We thank the KEKB group for the excellent operation of the
accelerator, the KEK cryogenics group for the efficient
operation of the solenoid, and the KEK computer group and
the National Institute of Informatics for valuable computing
and Super-SINET network support. We acknowledge support from
the Ministry of Education, Culture, Sports, Science, and
Technology of Japan and the Japan Society for the Promotion
of Science; the Australian Research Council and the
Australian Department of Education, Science and Training;
the National Science Foundation of China and the Knowledge
Innovation Program of the Chinese Academy of Sciences under
contract No.~10575109 and IHEP-U-503; the Department of
Science and Technology of India; 
the BK21 program of the Ministry of Education of Korea, 
the CHEP SRC program and Basic Research program 
(grant No.~R01-2005-000-10089-0) of the Korea Science and
Engineering Foundation, and the Pure Basic Research Group 
program of the Korea Research Foundation; 
the Polish State Committee for Scientific Research; 
the Ministry of Education and Science of the Russian
Federation and the Russian Federal Agency for Atomic Energy;
the Slovenian Research Agency;  the Swiss
National Science Foundation; the National Science Council
and the Ministry of Education of Taiwan; and the U.S.\
Department of Energy.

\end{document}